\documentclass[journal,twoside,web]{IEEEtran}
\usepackage{cite}
\usepackage{amsmath,amssymb,amsfonts}
\usepackage{algorithmic}
\usepackage{graphicx}
\usepackage{textcomp}
\usepackage{hyperref}
\usepackage{fancyhdr}
\usepackage{graphicx}
\usepackage{multicol}
\usepackage{subfigure}
\usepackage{amsmath}
\usepackage{amssymb}
\usepackage{makecell} 
\usepackage{multirow}
\usepackage{booktabs,caption}
\usepackage{float}
\usepackage{lipsum}
\usepackage{dblfloatfix}
\usepackage{blindtext}
\usepackage{ragged2e}
\usepackage{xcolor}
\usepackage{array}
\usepackage{mathptmx} 
\usepackage{amsmath}
\usepackage{mathrsfs}
\usepackage{ragged2e}

\begin{document}
\title{Futuristic Variations and Analysis in Fundus Images Corresponding to Biological Traits}
\author{Muhammad Hassan, Hao Zhang, Ahmed Fateh Ameen,Home Wu Zeng,Shuye Ma, Wen Liang, Dingqi Shang, Jiaming Ding, Ziheng Zhan, Tsz Kwan Lam, Ming Xu, Qiming Huang, Dongmei Wu, Can Yang Zhang, Zhou You, Awiwu Ain{*}, and Pei Wu Qin{*}
\thanks{This research was supported by the National Natural Science Foundation of China (Grants Nos. 31970752); Science, Technology, Innovation Commission of Shenzhen Municipality (Grants Nos. JCYJ20190809180003689, JSGG20200225150707332, JSGG20191129110812708, ZDSYS20200820165400003, WDZC20200820173710001); and Shenzhen Bay Laboratory Open Funding (Grants No. SZBL2020090501004). }
\thanks{M. Hassan, Hao Zhang, S. Ma, W. Liang, D. Shang, J. Ding, Z. Zhan, T. Lam, M. Xu, Q. Huang, D. Wu, A. Ain and P. W. Qin are with Shenzhen Internatioanl Graduate School Tsinghua University (mhassandev@gmail.com,pwqin@sz.tsinghua.edu.cn). }
\thanks{M. Hassan, A. F. Ameed, H. Zeng are with Shenzhen Children's Hospital, Guangdong China.}
\thanks{Zhou You is in School of Computer Science and Technology, Jilin University}}

\maketitle

\begin{abstract}
Fundus image captures rear of an eye, and which has been studied for the diseases identification, classification, segmentation, generation, and biological traits association using handcrafted, conventional, and deep learning methods. In biological traits estimation, most of the studies have been carried out for the age prediction and gender classification with convincing results. However, the current study utilizes the cutting-edge deep learning (DL) algorithms to estimate biological traits in terms of age and gender together with associating traits to retinal visuals. For the trait’s association, our study embeds aging as the label information into the proposed DL model to learn knowledge about the effected regions with aging. Our proposed DL models, named FAG-Net and FGC-Net, correspondingly estimate biological traits (age and gender) and generates fundus images. FAG-Net can generate multiple variants of an input fundus image given a list of ages as conditions. Our study analyzes fundus images and their corresponding association with biological traits, and predicts of possible spreading of ocular disease on fundus images given age as condition to the generative model. Our proposed models outperform the randomly selected state-of-the-art DL models.

\end{abstract}


\section{Introduction}
\label{sec:introduction}
\IEEEPARstart{T}{he} retina is the organ that enables humans to capture visuals from the real world. The retina is a vital source to assess distinct pathological processes and neurological complications associated with risks of mortality. The retina refers to the inner surface of the eyeball opposite to the lens, including the optic disc, optic cup, macula, fovea, and blood vessel~\cite{5660089,navab2015medical}. Fundus images are fundus projections captured by a monocular camera on a 2D plane~\cite{li2021applications}. Fundus images play an important role in monitoring the health status of the human eye and multiple organs~\cite{bernardes2011}. Analyzing fundus images and their corresponding association with biological traits can help in preventing eye diseases and early diagnosis. Retina allows us to visualize both vascular and neural tissues in a non-invasive way and examine neurological complications. The strong association of retina with physiology and vitality may lead to a deeper association with biological traits, such as age and gender. Biological traits can be determined by genes, environmental factors, or a combination of both, which can either be qualitative (such as gender or skin color) or quantitative (such as age or blood pressure)~\cite{mackay2009genetics}. Biological traits are relevant to a variety of systemic and ocular diseases in individuals~\cite{betzler2021}, for instance, females are expected to have longer life expectancies compared to those males in similar living environment~\cite{austad2006women,zarulli2018women,baum2021new}. With the increasing age, women with reduced estrogen production are predisposed to develop degenerative eye diseases, including cataracts and age-related macular degeneration~\cite{klein1992prevalence,rudnicka2012age,rim2014cataract}. In contrast, males are more likely to suffer from pigment dispersion glaucoma~\cite{scheie1981pigment}, open-angle glaucoma~\cite{rudnicka2006variations}, and diabetic retinopathy~\cite{zhang2010prevalence}. The study of biological traits association with fundus images is a challenging task in the clinical practices where experts in the field are unaware of the gender discrimination in the fundus images of males and females and the association of aging information. Thus, this study utilizes deep learning (DL) algorithms to estimate biological traits and their association with the generated fundus images. 

The fundus or retinal images have been studied for classification, disease identification, and analysis using conventional machine learning (ML) to recent DL methods~\cite{poplin2018prediction,githinji2022multidimensional}. However, much of the work has been focused on ‘feature engineering’, which involves computing explicit features specified by experts. On the contrary, DL has been characterized by multiple computational layers that allow an algorithm to learn the appropriate predictive features based on examples. The DL algorithms have been utilized for the classification and detection of different eye diseases, such as diabetic retinopathy and melanoma, with human comparable results. In the conventional ML approaches, the relationship between retinal morphology and systemic health has been extrapolated using multivariable regression. However, such methods show limited ability for large size and complex datasets~\cite{mutlu2018association,owen2019retinal}. Thus, the DL algorithms avoid manual feature engineering, tuning, and made it possible of extracting hidden features which were previously unexplored by the conventional methods. The DL models have shown significant results for previous challenging tasks. The harnessing of DL power innovatively associated the retinal structure and pathophysiology. The DL models can extract independent features unknown to clinicians; however, they may face challenges of explainability and interpretability, which have been attempted to address in the existing work~\cite{hassan2022neuro}. The DL approaches to fundus image analysis are receiving popularity featuring easy implementation and high efficiency~\cite{normando2016retina}. It has been extrapolated that DL models can capture subtle pixel-level information in terms of luminous and contrast which humans may not differentiate. These findings underscore the promising ability of DL models hidden to humans and can be employed in medical imaging with high efficacy in clinical practices~\cite{korot2021predicting}.

In clinical studies, experts in the field are unaware of the subjects' discrimination based on their fundus images which emphasis on the importance of employing DL models. The cause and effect of demographic information in fundus images are not readily apparent to domain experts. On the contrary, DL models may enable data-driven algorithms to discover of novel approaches to disease biomarkers identification and biological traits association. Therefore, the ophthalmoscope has been deeply associated with systemic indices of biological traits (such as aging and gender) and diseases. In previous studies, age has been estimated from distinct clinical images, such as age prediction from brain MRI, facial images, and neuroimaging using machine learning and deep learning~\cite{wang2019gray,xia2020three,cole2017predicting,jonsson2019brain}. For instance, brain MRI and facial images have been used for age prediction emphasizing on the potential of traits estimation from fundus images~\cite{wang2019gray,xia2020three,cole2017predicting,cole2018brain}. The excellent performance in age prediction implies that fast, safe, cost-effective, and user-friendly deep learning models can be possible in a larger population. In addition to the aging association, fundus images have also been associated with sex by applying logistic regression on several features~\cite{yamashita2020factors}. These features include papillomacular angle, retinal vessel angles, and retinal artery trajectory. Various studies have shown retinal morphology differences between the sexes, including retinal and choroidal thickness~\cite{ooto2015effects,lamparter2018association}. The study~\cite{korot2021predicting} reported fovea as an important region for gender classification. The prediction of gender became possible, which was an inconvenient job for the ophthalmologist who spent the whole career at retina~\cite{ting2018eyeing}. Thus, results for the age and gender estimation may assist investigating physiological variations on fundus images corresponding to biological traits~\cite{poplin2018prediction}. The estimation of age and gender classification may not be clinically inevitable, but the study of age progression based on biological traits learning hints the potential application of DL in discovering novel associations between traits and fundus images. The DL models implementation uncovers additional features from fundus images results in better biological traits association~\cite{khan2022predicting}.  

The successful estimation of age and gender prediction convince for studying age progression effects and evaluating aging status via fundus images. In the study of~\cite{poplin2018prediction}, aging effects were investigated while associating cardiovascular risk factor with fundus images. Similarly, large size DL models was used for classification and association of fundus images with physiological traits dependent on patients' health~\cite{poplin2018prediction}. The existing algorithms mainly consider the optic disc's features for gender prediction as having consistent observations with that of Poplin~\cite{poplin2018prediction}. In Poplin's work, large deep learning models were used to classify sex and other physiological and behavioral traits that were associated with patient health based on fundus images. Similarly, fundus (retinal) images were closely related to age and gender traits by allowing the definition of 'retinal age gap', which is a potential biomarker for aging and risk of mortality~\cite{zhu2022retinal}. 

The variational effects of age progression can be visualized in distinct ways, including saliency maps or heat maps in fundus images that were difficult to be observed by ophthalmologists. The differential visualization in fundus images can also be used to distinguish male and female subjects. After the successful classification of gender trait from fundus images~\cite{betzler2021gender}, our proposed model (FAG-Net) emphasizes on optic disc area and learned features while training and learning the association corresponding to aging. The optic disc was also considered the main structure to train our deep learning approaches. Similarly, the second proposed model (FGC-Net) utilizes the learned knowledge to generate different fundus images given a single fed fundus with a list of ages as labels (conditions). FGC-Net can be used to predict the possible spreading of eye diseases and early diagnosis. To carry out this, firstly, we trained and successfully evaluated a DL model (FAG-Net) for the trait effects in terms of age and gender estimation. Secondly, we proposed a DL model (FGC-Net) to learn aging effects and embed these effects for the generation of multiple fundus versions to highlight the possible disease spreading. The corresponding multiple generated versions are subtracted accordingly to demonstrate the learning effects with age progression and disease projections. The rest of the manuscript is organized as follows: Section-2 outlines the existing works, Section-3 demonstrates methods, Section-4 illustrates and analyzes results, and Section-5 concludes the study with future directions.

 \section{Literature study}
In previous studies, age and gender have been estimated from distinct imaging modalities such as age prediction from brain MRI, facial images, and neuroimaging using machine learning and deep learning~\cite{wang2019gray,xia2020three,cole2017predicting,jonsson2019brain}. Brain MRI and facial images have been used for age prediction emphasizing on the potential of traits estimation from fundus images~\cite{wang2019gray,xia2020three,cole2017predicting,cole2018brain}. In the work by Poplin~\cite{poplin2018prediction}, large deep learning models were used to classify sex and other physiological and behavioral traits that were associated with patient health based on retinal fundus images. There are a number of studies in which the fundus images have been used for age prediction and gender classification using machine learning classification~\cite{korot2021predicting,yamashita2020factors,ooto2015effects,lamparter2018association,wang2019gray,xia2020three,cole2017predicting,cole2018brain}. Most of them have estimated the age and gender either from healthy or unhealthy subjects. However, the current study underlying both healthy and unhealthy subjects for age and gender association to fundus images. For age and gender prediction, conventional to recent deep learning based algorithms have been employed~\cite{normando2016retina,poplin2018prediction,korot2021predicting,zhangai}. To our knowledge, non-of them attempted the age progression effects besides the age prediction and gender classification. 

Clinicians are currently unaware of the distinct retinal feature varying between males and females, highlighting the importance of deep learning and model explainability. Automated machine learning (AutoML) may enable clinician-driven automated discovery of novel insights and disease biomarkers. Gender has been classified in the study of~\cite{korot2021predicting}, in which, the area under the receiver operating characteristic curve of the code free deep learning model was 0.93. The study~\cite{liu2019biological} estimated biological age from dataset collected with MAE = 3.67 and cumulative score = 0.39~\cite{jin2018prevalence}. The study~\cite{kim2020} developed CNN age and sex prediction models from normal participants without hypertension, diabetes mellitus (DM), and any smoking history. However, our proposed model (FAG-Net) shows higher results in the majority of the evaluation metrics compared to the existing models for both healthy and unhealthy subjects as tabulated in Table~\ref{tab:age-prediction-FAG-Net-vs-SOTA} and \ref{tab:Gender-Classification-FAG-Net-vs-SOTA}.

The ML algorithms are widely applied in analyzing biological traits with different imaging modalities. In the conventional biological traits estimation, the study~\cite{jia2020tsea} proposed a trait tissue association mapping human biological traits and diseases. The study~\cite{franke2010estimating} estimated the age of the subjects using PCA~\cite{abdi2010principal} for dimension reduction and relevance vector machine~\cite{tipping1999relevance} with significant score. Similarly, study~\cite{valizadeh2017} used neural network and support vector machine to analyze five anatomical features, which resulted in high prediction accuracy. According to the study of~\cite{poplin2018prediction}, machine learning has been leveraged for many years for a variety of classification tasks, including the automated classification of eye disease. However, much of the work has focused on ‘feature engineering’, which involves computing explicit features specified by experts. 

The relationship between retinal morphology and systemic health has been extrapolated using multivariable regression like conventional approaches. However, such methods show limited ability for large size and complex datasets~\cite{mutlu2018association,owen2019retinal}. Thus, the advances in automatic algorithm into DL avoids manual feature engineering and made extracting hidden features possible, which were previously unexplored. The DL models have shown significant results for previous challenging tasks. The harnessing of DL power is innovatively associated with the retinal structure and pathophysiology. DL models extract independent features unknown to clinicians; however, face challenges of explainability and interpretability which have been attempted to address by a neuro-symbolic learning study~\cite{hassan2022neuro}. Deep learning has been applied in different domains specifically in diseases diagnosing, such as melanoma and diabetic retinopathy, and achieved comparable accuracy to that of human experts~\cite{zhang2022rcmnet}. 

Deep learning approaches to automated retinal image analysis are gaining popularity for their relative ease of implementation and high efficacy~\cite{normando2016retina}. It has been reported that DL models capture subtle pixel-level luminance variations, which are likely indifferentiable to humans. Such findings underscore the promising ability of deep neural networks to utilize salient features in medical imaging which may remain hidden to human experts~\cite{korot2021predicting}. Deep learning has shown great strength in medical image analysis. Most importantly, ophthalmologists have successfully predicted biological traits, such as age and gender with the significance of 0.97 as area under curve (AUC) score~\cite{korot2021predicting}. Yamashita performed logistic regression on several features that were identified to be associated with sex~\cite{yamashita2020factors}. These features include papillomacular angle, retinal vessel angles and retinal artery trajectory. Various studies have shown retinal morphology differences between the sexes, including retinal and choroidal thickness~\cite{ooto2015effects,lamparter2018association}. In previous studies, age has been estimated from clinical images via machine learning and deep learning~\cite{wang2019gray,xia2020three,cole2017predicting,jonsson2019brain}. The excellent performance in age prediction implies that fast, safe, cost-effective, and user-friendly deep learning models are possible in larger population. In comparison to the state-of-the-art works, our study aims to show the continuous effect of age progression besides age estimation and gender classification.

\section{Methodology} 
\subsection{Biological Traits Estimation using FAG-Net Architecture}
\begin{figure}[h]
	\centering
	\includegraphics[width=1\linewidth]{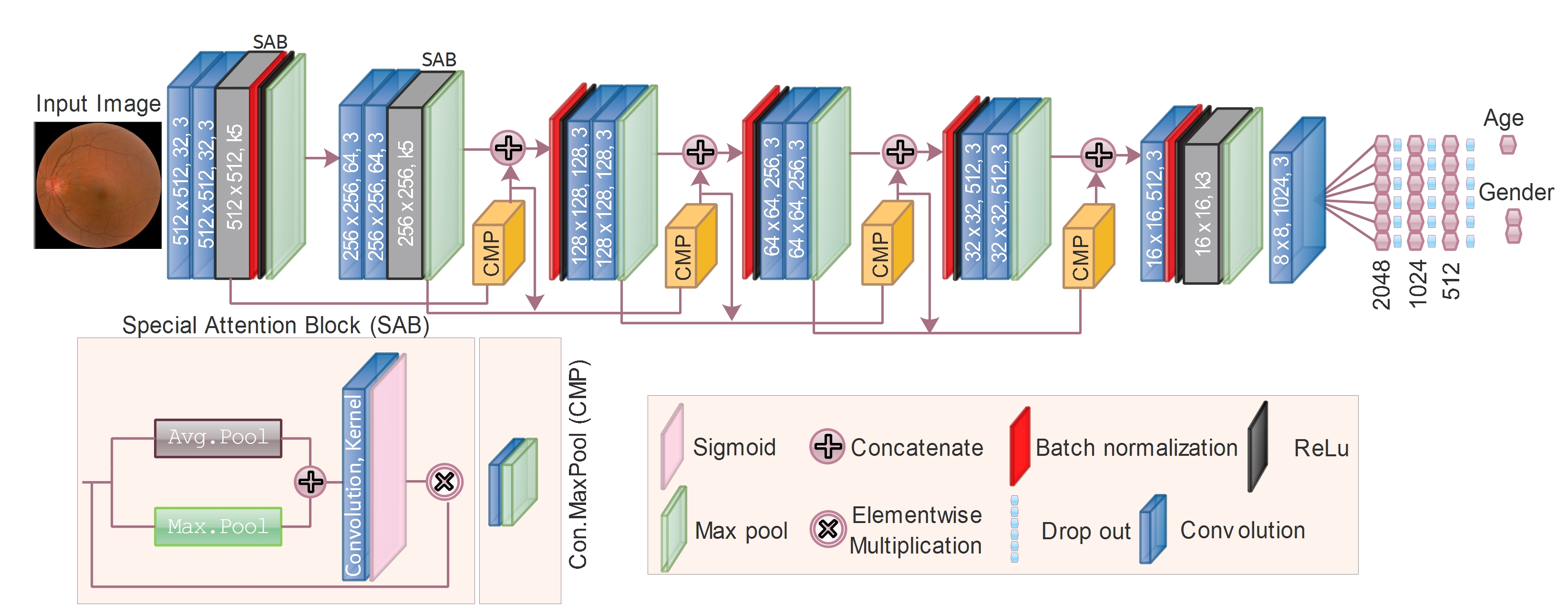}
	\caption{FAG-Net: for age and gender estimation using fundus images. }
	\label{fig:figure1}
\end{figure}

For Age prediction and gender classification, we borrowed the concept of biological traits estimation from ShoeNet model~\cite{hassan2021deep}. The ShoeNet model has been used for age estimation and gender classification from pairwise shoeprints. However, the datasets available for fundus images are rarely found in pairwise (left and right eyes' images). Thus, the model needs special attention to overcome the challenging situation to be utilized for biological traits estimation. Therefore, we propose a model for \textbf{f}undus images based \textbf{a}ge and \textbf{g}ender estimation (FAG-Net) (Figure~\ref{fig:figure1}). The model composed of six blocks, where block1, -2, and -6 contain spatial attention mechanism (SAB) while the rest of the blocks have been exempted from SAB. The first block receives input fundus images with dimensions of 512$\times$512$\times$3 (width$\times$hight$\times$channel). The input three channels fundus image first pass through a stack of convolution neural network with given number of filters (32) and kernel size (3). The SAB block has been augmented to focus on the salient spatial regions. 

Attention-mechanism has shown great attention recently due to its significant performance in the literature~\cite{oktay2018attention}. In practice, both channel wise (CA) and spatial wise (SA) attentions have been employed with channel first order. However, we only applied SA, which only focuses along the spatial dimension. In SA, average pooling and max pooling are applied in parallel to the input and concatenated correspondingly. A 2D attention map is generated over each pixel for all spatial locations with a large filter-size (i.e., $K=5$). The convolutional output is then normalized by non-linear sigmoid function. Finally, the normalized and direct connections are merged with element-wise multiplication to produce attention-based output. Both average and max pooling are used in SA to balance the selection of salient features (max pool) and global statistics (average pool). The embedding of the attention mechanism in FAG-Net focuses on regions of interest vulnerable to aging effects. The output from SAB passes through batch-normalization (BN) and rectified linear unit (ReLu) functions. Therefore, each block ends with BN and ReLu functions. 
The input of block-2 received from block-1 passes through stack of convolution, SAB, and ends with maxpool layer. Similarly, the output of block-1 also passes as a direct connection to convolution maxpool (CMP) block. The convolution layer in CMP applies to the output (from block-1) with the same number of 64 filters (as that of block-2). However, the 1$x$1 kernel size has been used to produce the same feature maps (64) and passed to a maxpool operation to bring into the same dimension. Both outputs of block-2 and CMP block concatenate along the third dimension and forward as input to block-3 and a direct connection. 

The purpose of CMP block is to retain the spatial features in high dimensional space  to deeper-level related to age progression. In the abstract level, the dense structure passes salient features together with those extracted from block sequence. The feature maps increase and the dimensions decrease as the network goes deeper. The accumulated output from all the blocks passes through a normal convolution layer having 8$x$8 feature maps and 1024 number of filters. The final convolution layer passes the output to fully connected layers where each layer has been dropout with ratio of 9, 8, and 5 output 10 to avoid overfitting. The final output neuron can be singleton for age prediction or two for gender classification. In the case of age prediction, a linear activation function applies to produce a regression value. However, for gender classification, a softmax layer is employed to output weighted output for both male and female.  

\subsection{Objective function for FAG-Net}
The objective function used for training FAG-Net composed of three loss terms including $L_1$, $L_2$, and regression specific custom loss function (CLF). The accumulative loss function (ALF) is the mean of all the weighted loss terms, formulated as follows:
\begin{equation}
ALF = \psi*L_1+\psi*L_2+\psi*CLF)/3,
\end{equation}
$\psi$ is the corresponding weights to balance the loss terms, including $L_1$ and $L_2$ and which can be formulated as follows:
\begin{equation}
\textit{L}_\textit{1} =  {\sum_{i=0}^{n-1} abs|A_{age}^i-P_{age}^i|}\label{equ:NAE},
\end{equation}
\vspace{-0.15cm}
\begin{equation}
\textit{L}_\textit{2} =  {\sum_{i=0}^{n-1}\{A_{age}^i-P_{age}^i\}^2}\label{equ:MSE},
\end{equation}
where $n$ is the number of samples and $A_{age}$ and $P_{age}$ denote the actual and predicted ages. 

Furthermore, age prediction is a regression problem, and a single output will be expected as a result. Thus, a specialized custom loss function based on mean square error (MSE) is proposed to optimize the hyperparameters during training~\cite{hassan2021deep}. The optimizer (Adam) fine-tunes the weights of convolution filters to minimize the loss value. To produce regression specific results, CLF penalizes the out-ranged values more. It minimizes the distance between the actual and predicted age in a target-oriented way. The formulation of CLF is illustrated in the following equation:		
\begin{equation}
\textrm{CLF}=
\frac {\sum_{i=1}^{n} E_i}{n};~E_i=
\begin{cases}
d_i*\varphi,& \text{if	}  d_i\leq J\\
d_i^{3}+\varphi,& \text{if	}  d_i>J
\end{cases}\label{equ:2}
\end{equation}
CLF is the mean of difference $(E)$ for $n$ number of samples, where $n$ = $(total-samples)/(input$-$size)$. $\varphi$ is a small value (0.0001-to-0.3) used to prevent the network from attaining zero difference and to sustain the learning process. Similarly, $d_i=||y-\bar{y}||$ is an absolute error between actual age $(y)$ and predicted age $(\bar{y})$. Furthermore, $J$ is a natural number derived from \textit{MCS-J} for predictable age ranges. In the second condition ($d_i\leq J$), the values higher than the value of $J$ will cause more penalization of the weights based on the computed loss-value in the exponential time (power 3). The penalization influences the optimization of network weights and biases. It will direct the optimizer to tune these parameters in order to minimize the difference between actual and predicted age. CLF not only considers the absolute error but also penalizes more the adjacent values to $J$ in MCS-J. Adam is used as an optimizer with the L$_2$ regularizer to tune hyper-parameters following the objective function.

\subsubsection{Evaluation Metrics for FAG-Net} 
Besides MAE and MSE as evaluation metrics for age prediction, we apply cumulative score (CS) and mean cumulative score (MCS) as evaluation metrics to accommodate the nature of the problem. CS and MCS imitate the existing studies, and are used to assess accuracies in a range of age groups. CS (or CS$_j$) and MCS (or MCS-J) give more weight to the smaller ranges of match windows. The ranges depend on the value of $j$ and $J$, the absolute differences between actual and estimated age scores~\cite{hassan2021deep}, is formulated as follows.
\begin{equation}
\textrm{MCS-J}=\frac {\sum_{j=0}^{J} CS_j}{J+1}\label{equ:MCS-J},~
\textrm{CS$_j$}=
\frac {\sum_{i=1}^{n} \delta_i}{n}*100,\\~
\delta_i=
\begin{cases}
1,& \text{if	}  \delta_i\leq j\\
0,& \text{if	}  \delta_i>j\\
\end{cases}
\end{equation}
CS$_j$ is the percentage mean of $\delta_i$, where $\delta_i$ is the Euclidean-distance $(|y_i-\bar{y_i}|)$ between actual ($y_i$) and predicted ($\bar{y_i}$) score, and it will be counted as 1 for $|y_i-\bar{y_i}|\leq j$. The value of $\delta_i$ expressed as zero (0) implies that the distance $(|y_i-\bar{y_i}|)$ is greater than the threshold value ($j$). The MCS score facilitates prediction in various ranges of matching thresholds rather than a single threshold. Thus, the MCS score gives a more comprehensive assessment for the challenging problem of retinal based age prediction to cover all the values of $|y_i-\bar{y_i}|\leq j$ for the setup threshold ($j$). This also allows us to give different penalties with varying thresholds in the objective function of the deep learning model.

\subsection{Fundus Images Generation Given Age as Condition}
\begin{figure}[h!]
	\centering
	\includegraphics[width=1\linewidth]{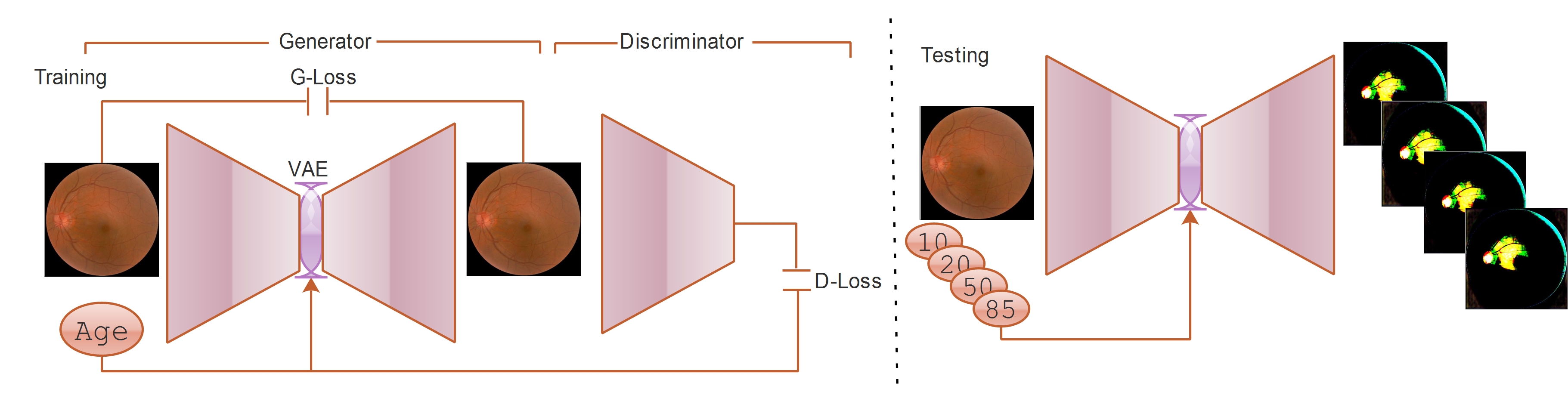}
	\caption{FGC-Net: for fundus images generation given different ages. The generator part composed of variational autoencoder where the bottleneck receives age as condition prior to decoding. The discriminator receives both ground truth and generated fundus images to learn the aging effects. \textbf{Left}: The FGC-Net receives fundus image as input, encodes into latent space and generated back with embedded (in the bottleneck) age as condition. The generated fundus image discriminates against the age as label for learning age embedding. \textbf{Right}: for the testing phase, a single fundus image can be input with multiple age labels and generates corresponding fundus images with relevant variations. The details of the model have been drawn in Figure~\ref{fig:figure3}. }
	\label{fig:figure2}
\end{figure}

After proposing a sophisticated DL model (FAG-Net) for age prediction and gender classification, a novel network model has been introduced to predict futuristic variations in the fundus images. The model \textbf{g}enerates \textbf{f}undus images given age as \textbf{c}ondition (FGC-Net) (Figure~\ref{fig:figure2}). The FGC-Net 
\subsubsection{Encoding}
\begin{figure}[h!]
	\centering
	\includegraphics[width=01\linewidth]{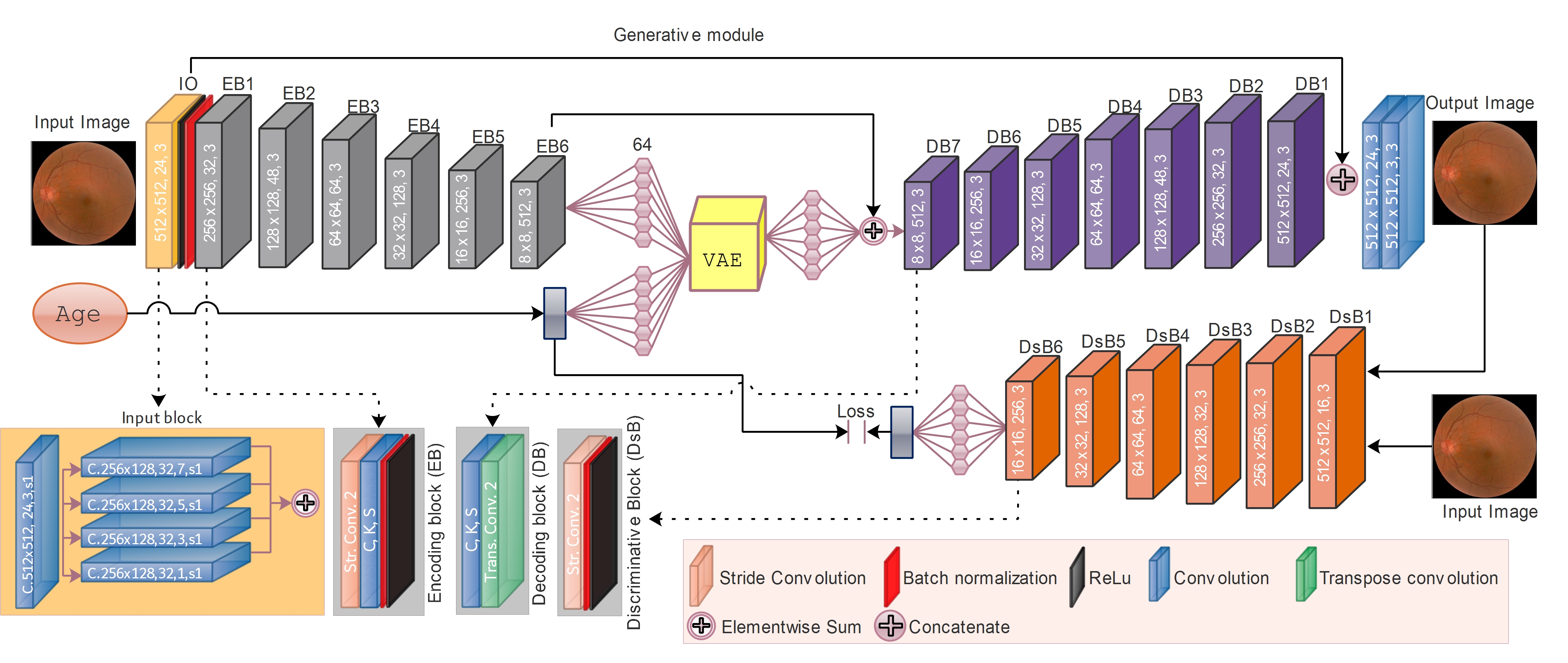}
	\caption{Detailed network architecture of FGC-Net, which is composed of generative and discriminative modules together with condition in the bottleneck using variational autoencoder (VAE). The VAE in the bottleneck embeds age as scaler values and facilitates different versions for the given input. The input block has a special architecture where varieties of filters (with different sizes) have been employed. The model output different copies of the input given ages as condition while testing. }
	\label{fig:figure3}
\end{figure}

The encoding phase of FGC-Net first receives the input fundus images ($X^i \mathbb{R}^{N\times H\times W\times C}$) regarding biological traits association and learning (Figure~\ref{fig:figure3}).  The dimensions $N\times H\times W\times C$, denote the batch size, width, height, and features map (number of channel: 1 for grayscale and 3 for color images), respectively. The encoder automatically extracts lower-dimensional features from the input data and inputs them into the latent space. The $i^{th}$ convolutional layer ($NC_i$) acts as a feature extractor by encoding the salient features from $X_i$. Considering the input structure (e.g., $X^h=H$, $X^w= W$, $X^c= C$, where $X^h$, $X^w$, $X^c$ are the output structure with new height $h$, width $w$, and dimension $c$, respectively), the encoder (e) part contains five encoding blocks-EB  (EB-1 to -6) to sufficiently extract low-level features in the spatial dimension (e.g., $X^h=\frac{1}{n}\times H$, $X^w=\frac{1}{n}\times W, X^c = n\times C$, where $n$ is the number of downsampling and deeper levels) followed by the bottleneck layer ($Z\in \mathbb{R}^k$, where k is the spatial dimension of $Z$). The size of the channels (EB-1 to EB-6) decreases by halves in each subsequent deep layer, where the loss of information can be compensated for by doubling the number of filters (channels). 
In the encoding layer, the received image passes through an input-block (IB) which has been designed for the purpose of extracting varieties of features by employing distinct kernel and filter sizes such as 1$\times$1, 3$\times$3, 5$\times$5, and 7$\times$7 after a normal convolution (512$\times$512, 24, 3, corresponding to dimensions, number of filters, and kernel size). The outputs of variant size of filters merge as elementwise sum prior to proceeding for the deeper block. The output of IB forwards to EB1, where EB1 contains strides convolution to avoid the loss of information useful for generation, normal convolution (dimensions, number of features, kernel size, and stride) for feature extraction, followed by BN and ReLu functions. The rest of the blocks (EB2 to EB6) have the same structure till the bottleneck layer. The EB compresses the input spatial wise and extends channel wise. The compressing process at the $l^{th}$ encoding block $EB\mbox{-}l$, where $l=6$, is formulated as follows:
\begin{equation}
EB\mbox{-}l = En\Big(\left[ NC[S_t(X^{l-1})];\{op_b, op_r\} \right];\phi\Big)\label{equ:ecoding-layer},
\end{equation}
where, $S_t$ and  $Co$ denote strides (s = 2) and normal convolution in a block ($-l$)  over the data sample ($X^{l-1}$) obtained from a previous block ($l-1$). The output from a stride convolution $S_t$ and normal convolution $NC$ forwards to BN ($op_b$) and ReLu ($op_r$) functions. The stack of stride convolution ($St$) and normal convolution ($NC$) avoids the loss of useful information. In addition to reducing computational operations~\cite{he2016deep}, $St$ enables the model to learn while downsampling~\cite{ayachi2018strided} and retain and passes features into subsequent layers heading into latent space which is used by the decoder to generate back with age embedded effects. Besides the encoding of input fundus image, the corresponding label information has also been embedded into the latent space. After a number of experiments, the label information as condition in the latent space is more effective to effect the generation process. The embedding of age as condition with the output of the encoding layer is carried out in the latent space. The encoder part ($En$) passes the label (age $L_g$) information as condition ($Ec(L_g, \xi)$), where $\xi$ is the learned parameters by the encoder, into the latent space of VAE.  

\subsubsection{Bottleneck and Conditioning}
The bottleneck layer is an unsupervised method of modeling complex and higher dimensional data in deeper level. The encoder part ($En(X^i, \phi)$) compresses the input from higher-dimensional space ($X_m^H,X_m^W$) through network parameters ($\phi$) and generates the probabilistic distribution over the latent space ($Z$) with a lower possible dimension ($\frac{1}{n}\times X^H,\frac{1}{n}\times X^W$). Similarly, $Ec(L_g, \xi)$ passes $L_g$ through fully connected layers with learning parameters of $\xi$ similar to the fully connected layer of $En(X^i, \phi)$. The decoder part utilizes the embedded and compressed form (latent variables $Z$) and generates it back to the high-dimensional generated space (${Y}$). Minimizing the gap between $X$ and $Y$ enables the model to learn and tune the parameter values. The latent space enables the model to learn from the mutual distributions of $X$ and $L_g$. The output of $EB6$ and $En(L_g, \xi)$ are passed to a fully connected layer modeling the complex dimensional structure into a latent representation and then flatten via 64 neurons. From each of the flatten 64-neurons, both mean ($\mu$) and standard deviation ($\sigma$) are computed.

The encoder part ($En\{X_{m};~\phi\}$) generates the posterior over latent space ($z^i$, where $i$ denotes the sample number) and samples from the latent space ($P^i$) which can be used for the decoding (generation) as $De\{En\uplus L_g \oplus S_f;~\vartheta\}$. The latent space is obtained as follows:
\begin{equation}\label{Equa:latent-space}
\begin{split}
z_i\sim \Re_i \left[(z_0|x^i)\parallel (z_1|x^i) \right],
\end{split}  
\end{equation}
where $\Re(|)$ is the distribution over $z_0$ and $z_1$ given input $x^i$. The sampling $z_i$ from the distribution $\mathcal{N}(;)$ can be rewritten for the conditional input as follows: 
\begin{equation*}\label{equ:latent-distribution}
\begin{split}
z_i\sim \mathcal{P}_i(z|X)& = \mathcal{N}\big(\mu(X; \phi_0), \sigma(X; \phi_0)\big)\parallel \mathcal{N}\big(\mu(X; \phi_1), \sigma(X; \phi_1),\\
z_i\sim \mathcal{P}_i(z|X)& = \mathcal{N}\big( \left[ \mu(X; \phi_0)+\mu(X; \phi_1)\right]; \left[ \sigma(X; \phi_0)*\sigma(X; \phi_1) \right] \big),\\
z_i\sim \mathcal{P}_i(z|X)& = \mathcal{N}\big( \left[ \mu(X; \phi_l)\right]; \left[ \sigma(X; \phi_m) \right] \big),\\
where ~ \phi_l&=\phi_0+\phi_1, ~\phi_m=\phi_0*\phi_1
\end{split}
\end{equation*} 
The drawn sample ($z_i$) conditioned with $X_m$ from the distribution (see~(Equation~\ref{equ:latent-distribution})) maps into the same shape as the decoder ($Dec(z_i, \theta)$) for the generation process with the learning network parameters ($\theta$). The latent distribution must be regularized by the kullback leibler (KL) divergence (see the loss function) to closely approximate the posterior ($P(z|x)$) and prior ($P(z)$) distributions. The regularization (i.e., via the Gaussian prior) holds in the latent space between the distributions in terms of $\mu$ and $\sigma$, which further contributes to the latent activations utilized by the decoder to produce new retinal image. The latent distributions are centered ($\mu$) and spread over the area $\sigma$ to project the possible fundus as desired (DSp). Usually, the distance between the learned distribution $\mathcal{N}(\mu,~\sigma^2)$ and the standard normal distribution $\mathcal{N}(0,1)$ can be quantified by the KL divergence. However, instead of Gaussian normal distribution and normal mean ($\mu$) standard deviation $\sigma$, we utilize the sum of $\mu$ and the product of $\sigma$. The detailed formulation is shown in Equations~\ref{equ:latent-distribution} and~\ref{equ:Z-generation}.  The latent distribution and regularization are expected to have the properties of continuity and completeness. In the case of continuity, the sampling from the latent distribution given $X$ may exist a nearby data point that feeds into the decoder to generate fundus images with a similar structure with additional information, as desired. The decoder must generate target-oriented fundus images in a controlled fashion.

\subsubsection{Decoding}
FGC-Net generates a random sample ($z_i, for~i= 1, 2 ,..., n$ ) conditioned by $L_g$ drawn from the probabilistic distribution $P_i(z_i|X)$ at the decoding side as decoding blocks (DB1 to DB7) and projects to $Y_i$:	
\begin{equation}
Y_i=Dec\big\{[z_i\odot R_i]\oplus S_f(X); \theta\big\}\label{equ:decoding-layer},
\end{equation}
where, $Y_i$ is the generated fundus images corresponding to $z_i$ with adjustable weights ($\odot R_i$) regularized by the objective function and merged with the contextual skipped features ($\oplus S_f$) using network learning parameters ($\theta$).  

In the decoding process, $z$ is computed from the sum of $\mu$ and $\sigma$ multiplied by the standard normal distribution ($\varepsilon$). The values of $\mu$ and $\sigma$ are computed in Equation~\ref{equ:latent-distribution}. The $\varepsilon$ value is computed from the absolute different of normal distribution having mean $\mu(L_g)$ and standard deviation $\sigma(L_g)$ based on the fed scaler age values. 
\begin{equation}\label{equ:Z-generation}
z\sim \mathcal{N}(\mu,\sigma^2)\cdot\epsilon\leftarrow \mu+\sigma^2\cdot\mathcal{N}(\mu(L_g), \sigma(L_g)).
\end{equation}
The dimension of $z$ is reshaped and upsampled to match the dimension of the corresponding encoding layer (EB6) and merge $\oplus S_f$ as elementwise sum with the skip layer from EB6. Each block receives input, performs upscale dimensions via transpose convolution followed by BN and ReLu function. Each decoding block-DB composed of strides convolution, BN, and ReLu activation. The output of DB1 concatenates with the skip connection from IB. 

\subsubsection{Skip Layer}The deeper the network, the more chances of losing key features due to the application of downsampling operations and the vanishing gradient problem~\cite{drozdzal2016importance}.  Similarly, to avoid the loss of contextual information~\cite{cai2019multi}, we adopted skip connections between the encoding ($Enc_k\{X;~\phi\}$) and decoding ($Dec_k\{z\oslash S_f;~ \theta\}$) at particular layers($_K$) to transfer spatial features and global information in terms of the input image structure. The skip layers integrate the learned features from early levels, avoid degrading shallow stacked networks and overcome gradient information loss by retaining key features during training. These connections also improve the end-to-end mapping of training and achieve an effective role in a deeper network. The sole purpose of adopted skip connections is to facilitate the decoder to maintain the existing input structure while generating on the decoding side together with synthetic information to reflect age progression. The dimensions and merging position with the corresponding layers, both at the bottleneck and decoder layers, are show, in Figure~\ref{fig:figure3}. After generating $z$ given $P(z|X)$ from the encoder (see Equation~\ref{equ:ecoding-layer}), the decoder part merges the data sample information from the latent space, conditioning information ($L_g$), and skip connection at a particular layer ($_k$) is formulated as follows:
\begin{equation}
DB_k = Dec\big(NC[S_t(Y^{k+1});\{op_b, op_r\}] \oplus S_f(X);\theta\big), \label{equ:decoder}
\end{equation} where $Y_{k+1}$,  $\oplus S_f(X)$), and $\oplus$ denote the previous tensor, skipped features, and merging operation for the elementwise sum, respectively. Additionally, $op_b$ and $op_r$ denote BN and nonlinear activation ReLu operations, respectively. In addition to the completeness and continuity properties of the VAE, the involvement of skip connections borrowed from U-Net controls the generation process. 
\subsubsection{Discriminator}
The discriminative part borrowed from generative adversarial network (GAN) is appended at the end of FGC-Net, which brings sharpness and better quality to the generated images~\cite{radford2015unsupervised}. Adversarial learning plays a min-max game to distinguish the original and fake (generated or synthetic) images. FGC-Net brings the inferencing features to reason at the latent space and generates fundus images as desired~\cite{larsen2016autoencoding}. However, instead of training in a min-max fashion, we utilize the discriminative part solely for prediction of a scaler value or regression similar to subjects' ages. There are six blocks receiving both input and generated fundus images. Each discriminative block (DsB) composed of stride convolution, BN, and ReLu functions. The output of stacked DsB ends with three fully connected layers containing 512, 256, and 128 neurons followed by dropout layers with ratio of 0.8, 0.7, and 0.6, respectively. Finally, the output of fully connected layer passes through linear activation function to a single neuron for age estimation. In the objective function, both outputs as single values (from input fundus and generated fundus images) are formulated as mean square error (MSE) or $L_2$ loss term. 

In our case, the generator maps $X_i$ to $Y_i^j$ where $j= L_g$ and $L_g$ can be age values, and the discriminator distinguishes $X_i$ and $Y_i^j$ as real and fake, respectively. The min-max game of learning in GAN can be formulated as follows:
\begin{flalign}
V(D,G) = \underset{G}{min}~\underset{D}{max}(D_{XY}, G_X)\label{Equation: 5-minmax},
\end{flalign}
Similarly, the generative ($G_X$) and discriminative ($D_{XY}$) operations can be illustrated in mathematical forms as follows:
\begin{flalign}
G_{X} = G\{\underbrace{En(X_i;~\phi)\rightarrow Y_i\sim Dec(Z_{i};~\theta)}_{Generative~Unit}\rightarrow \underbrace{Disc(\left[X_{i}, Y_i\right];~\Phi)}_{Discriminative~Unit}\}\\ \nonumber 
G_{X} = G(X_{i},~Y_{i};~\omega)~where~\omega=\{\phi,\theta, \Phi\}
\end{flalign}    
The discriminator plays a vital role in the abstract reconstruction error in the circumstances where VAE is infused in the network model. The discriminator part measures the sample similarity~\cite{larsen2016autoencoding} at both element and feature levels. In addition, the discriminator is made stronger to distinguish between real and fake images by including $L_2$ loss term. 
\subsubsection{Objective function for FGC-Net} 
The objective loss function for FGC-Net is composed of reconstruction loss ($L_2$) (Equation~\ref{equ:MSE}) and KL divergence loss~\cite{KLdivergence1997Loss}. The probabilistic distribution in VAE as inferencing model ($q_\phi(z|x)$) approximates the posterior (true) distribution ($p_\theta(z|x)$) in terms of KL-divergence to minimize the gap as follows~\cite{kingma2019introduction}:
\begin{equation}
\textit{KL}_{d}(q_\phi(z|x)||p_\theta(z|x)))=\mathbb{E}_{q_\phi}\Big[log \frac{q_\phi(z|x)}{p_\theta(z|x)} \Big],
\end{equation}  
In our case, the KL-divergence between the distribution $\mathcal{N}(\mu_i, \sigma_i)$ of the inference model with mean $\mu_i$ and variance $\sigma_i$, and the standard normal distribution $\mu(L_g), \sigma(L_g)$ (Equation~\ref{equ:Z-generation}) with mean $\mu$ and unit variance $\sigma$ can be formulated after the Bayesian inference simplification~\cite{duchi2007derivations} as follows:
\begin{equation}
\textit{KL}_{d}(\mathcal{N}(\mu, \sigma)||\mathcal{N}(\mu(L_g), \sigma(L_g))) =\frac{1}{2}\sum_{i=1}^{l}\big(\sigma_i^2+\mu_i^2-1-exp(\sigma_i^2), \big)\label{equ:kl-divergence-as-loss}
\end{equation} 
Thus, the total loss function for FGC-Net (TLF-FGC) composed of the following terms:
\begin{equation}
TLF{-}FGC =  (L_1+L_2+KL_d)/3\label{equ:TLF-FGC-loss}
\end{equation} 

%-------------------------------------------------------------------	
\subsubsection{Dataset preparation and network training} To train, evaluate, and test proposed models for biological traits estimation and traits based futuristic analysis, we used dataset Ocular Disease Intelligent Recognition (ODIR-5K)~\cite{PeakingUniversityODIR5k}, PAPILA~\cite{kovalyk2022papila}, and a longitudinal population based on 10-year progression collections (10Y-PC)~\cite{yan2018ten}. To propose a generalized model for biological traits estimation, we utilized both cross-sectional and longitudinal population. Furthermore, to cover the biological traits estimation, both healthy and unhealthy subjects were included so that the underlying DL model should learn features invariant to abnormalities. Similarly, varieties of cameras and environments have been used to capture different qualities of images modeling sophisticated DL networks. 
Both FAG-Net and FGC-Net have been trained via Adam for optimizing network parameters. For FAG-Net and FGC-Net, Adam optimizer was used with initial learning rate of 0.001, $\beta_1$ = 0.9, $\beta_2$ = 0.999, where the learning rate was decreased by $\frac{1}{10}$ after every 50 epochs. The batch size was composed of 16 samples according to the available GPU's memory size. The models run for 500 epochs and apply early stopping on observing poor result after each epoch.
\begin{table*}[h!]
	\caption{FAG-Net scores for five cross validation $CS_0$, $CS_1$, $CS_2$, $CS_3$, MAE, MSE, MCS.}
	\label{tab:age-prediction-FAG-Net-scores}
	\footnotesize		
	\centering
	\renewcommand{\arraystretch}{1}
	\begin{tabular}[h!]{l|c|c|c|c|c|c|c|c|c|c|c}
		\hline Network&MAE&MSE	 & $CS_0$ & $CS_1$ & $CS_2$&$CS_3$&$CS_4$&$CS_5$&MCS-2&MCS-3&MCS-4\\ \hline
		FCV-1&2.269	&22.026&32.736&69.263&80.133&84.132&85.756&87.172&60.710&66.566&71.174\\ 
		FCV-2&2.286&24.734&33.944&71.470&80.258&83.632&86.006&88.172&61.890&67.326&71.062\\
		FCV-3&2.286&25.573&79.921&81.842&83.239&84.941&86.905&88.256&81.667&82.485&83.369\\
		FCV-4&1.401&3.324&18.280&62.354&88.147&95.159&97.663&99.249&56.260&65.984&72.320\\
		FCV-5&0.517&3.961&83.282&93.147&94.718&95.258&96.377&97.76&90.382&91.608&92.562\\
		\textbf{Average}&\textbf{1.634}&\textbf{15.151}&\textbf{49.705}&\textbf{75.767}&\textbf{85.475}&\textbf{88.814}&\textbf{90.729}&\textbf{92.168}&\textbf{70.315}&\textbf{74.9400}&\textbf{78.098}\\
		
		\hline
	\end{tabular}
	\\
	\justify 5-fold cross validation (FCV) together with mean values have been illustrated. The values of MCS-2 and MCS-3 are based on $CS_0$ to $CS_5$ and formulated in Equation~\ref{equ:MCS-J}.
\end{table*}
\section{Results}
\subsection{Biological traits estimation}
The proposed model FAG-Net and the state-of-the-art (SOTA) models have been trained for 5-fold cross-validation (FCV). The evaluation metrics MAE, MSE, MCS-2 and MCS-3 were used for testing purposes. Evaluation metrics MCS can help better assessing the performance of the models for age prediction where age prediction can only be predicted in a range of values rather than classified value. Therefore, MSE metrics may produce larger value for larger difference between actual and predicted values in the case of outliers. Thus, MSE metric for such scenarios may not be a reliable option. The details of five cross validation have been shown in Table~\ref{tab:age-prediction-FAG-Net-scores}. Table~\ref{tab:age-prediction-FAG-Net-vs-SOTA} shows the results of all the underlying modalities corresponding to evaluation metrics. 
\begin{table*}[h!]
	\caption{Comparative evaluation-scores of FAG-Net and SOTA models in terms of MAE, MSE, MCS, and R$_2$. For the values of CS$_0$, CS$_1$, CS$_2$, and CS$_3$, see formulation in Equation~\ref{equ:MCS-J}. }
	\label{tab:age-prediction-FAG-Net-vs-SOTA}
	\setlength\tabcolsep{2pt} % let LaTeX compute intercolumn whitespace
	\footnotesize		
	\centering
	\renewcommand{\arraystretch}{1}
	\begin{tabular}[h!]{l|c|c|c|c|c|c|c|c|c|c|c|c}
		\hline Network&MAE&MSE & $CS_0$ & $CS_1$ & $CS_2$&$CS_3$&$CS_4$&$CS_5$&MCS-2&MCS-3&MCS-4&R$_2$\\ \hline
		AlexNet&2.788	&24.803&16.935&49.498&68.921&79.746&84.722&87.865&45.118&53.775&59.965&0.827\\ 
		VGG-Net 16 &5.430&65.727&5.805&19.030&33.260&44.172&55.608&63.771&19.365&25.567&31.575&0.543\\
		VGG-Net 19&104.830&2142.82&0.261&0.829&1.309&1.920&2.880&3.710&0.800&1.080&1.440&-147.795\\
		ShoeNet&4.754&54.343&8.031&23.439&38.716&51.942&62.199&71.322&23.395&30.532&36.866&0.622\\	
		Pixel RNN&34.080&1878.363&1.789&5.718&10.999&40.840&17.634&20.470&6.169&8.336&10.196&-12.407\\
		Highway Net&8.921&142.779&2.790&8.954&18.783&25.0321&34.860&41.857&10.176&13.890&18.084&0.008\\
		Residual Net&8.558&130.949&3.415&11.578&18.700&26.197&34.443&41.607&11.231&14.972&18.867&0.090\\
		Google Inception V4&2.740&29.056&21.169&56.612&74.508&82.191&85.202&87.254&50.763&58.620&63.937&0.798\\
		ResNeXt &7.233&97.577&5.761&15.539&24.268&33.566&42.034&49.367&15.189&19.783&24.233&0.322\\
		\textbf{FAG-Net}&\textbf{1.634}&\textbf{15.151}&\textbf{49.705}&\textbf{75.767}&\textbf{85.475}&\textbf{88.814}&\textbf{90.729}&\textbf{92.168}&\textbf{70.315}&\textbf{74.9400}&\textbf{78.098}&\textbf{0.889}\\
		
		\hline
	\end{tabular}
	\\
\end{table*}
\subsection{Gender classification}
In biological traits estimation, we also trained the proposed (FAG-Net) and few SOTA models. All the classification results are shown in Table~\ref{tab:Gender-Classification-FAG-Net-vs-SOTA}. After the successful classification of gender traits from fundus images, the proposed model (FGC-Net) emphasized more on optic disc area and learned features while training for aging association. In the study for gender classification~\cite{betzler2021gender}, optic disc was also considered the main structure by the deep learning approaches. 

To evaluate the performance of our proposed model for gender classification, we randomly chose few SOTA models and trained and tested them on the same dataset and parameters (Table~\ref{tab:Gender-Classification-FAG-Net-vs-SOTA}). We used confusion metrics to evaluate the results. The rest of the metrics include true positive (TP), false positive (FP), true negative (TN), false negative (FL), specificity, sensitivity, positive predictive value (PPV), negative predictive value (NPV), F$_1$ score, and accuracy. The derivation of these metrics is illustrated in the following equations. 
\begin{equation}\label{equ:preicssion-recall-f1-accuracy}
Sensitivity    = \frac{TP}{TP+FN},~\\
Specificity = \frac{TN}{TN+FP},\\
\end{equation} 
\vspace{-0.2cm}
\begin{equation}
F_1      = 2\times\frac{Specifity\times Sensitivity}{Specificity+Sensitivity},~\\ \nonumber
PPV = \frac{TP}{TP+FP},\\\nonumber
\end{equation}
\vspace{-0.2cm}
\begin{equation}\nonumber
NPV = \frac{TN}{TN+FN},~\\ \nonumber
Accuracy = \frac{TP+FN}{TP+TN+FP+FN}.\nonumber
\end{equation}
\vspace{-0.2cm}
\begin{equation}\label{equ:R2}
R_2 = 1 - \frac{RSS}{TSS},~
SSR= \sum_{i=1}^{j}(X_i-Y_i)^2,~
TSS =  \sum_{i}^{j}\big( (Avg(X_i^j)-Y_i)^2  \big)^2
\end{equation} where sum of square of residuals (SSR), total sum of square (TSS).
From the accumulated results, our proposed model FAG-Net outperforms the competitive SOTA models. VGG-Net-16 has received the second highest score in terms of accuracy. The convincing results of our proposed model encourage us to proceed with the age prediction from fundus images and learning the corresponding effect. 
\begin{footnotesize}
\begin{table}[h!]
	\caption{Comparative evaluation-scores of FAG-Net and SOTA models in terms of MAE, MSE, MCS.}
	\label{tab:Gender-Classification-FAG-Net-vs-SOTA}
	\setlength\tabcolsep{0.5pt} % let LaTeX compute intercolumn whitespace
	\footnotesize		
	\centering
	\renewcommand{\arraystretch}{1}
	\begin{tabular}[h!]{l|c|c|c|c|c|c|c|c|c|c}
		\hline Network&TP&FP&TN&FN&Specificity&Sensitivity&PPV&NPV&F$_1$&Acc\%\\ \hline
		AlexNet&662&552&408&778&0.58&0.61&0.54&0.65&0.60&60.3\\ 
		VGG-Net 16 &990&114&897&168&0.88&0.85&0.89&0.84&0.87&86.9\\
		VGG-Net 19&100&104&889&117&0.89&0.84&0.90&0.83&0.87&87.0\\
		ShoeNet&991&113&883&183&0.88&0.88&0.89&0.82&0.86&86.3\\	
		Pixel RNN&997&107&861&205&0.88&0.82&0.90&0.80&0.85&85.6\\
		Highway Net&673&541&483&703&0.56&0.58&0.55&0.59&0.57&57.3\\
		Residual Net&978&126&813&213&0.87&0.82&0.88&0.80&0.84&84.3\\
		
		\textbf{FAG-Net}&\textbf{1141}&\textbf{74}&\textbf{1065}&\textbf{121}&\textbf{0.935}&\textbf{0.904}&\textbf{0.939}&\textbf{0.897}&\textbf{0.919}&\textbf{91.87}\\
		\hline
	\end{tabular}
	\\
	\justify Positive Predictive Value (PPV), Negative Predictive Value (NPV)
\end{table}
\end{footnotesize}

\subsection{Age progression effects in fundus images}
\begin{figure}[h!]
	\centering
	\includegraphics[width=0.99\linewidth]{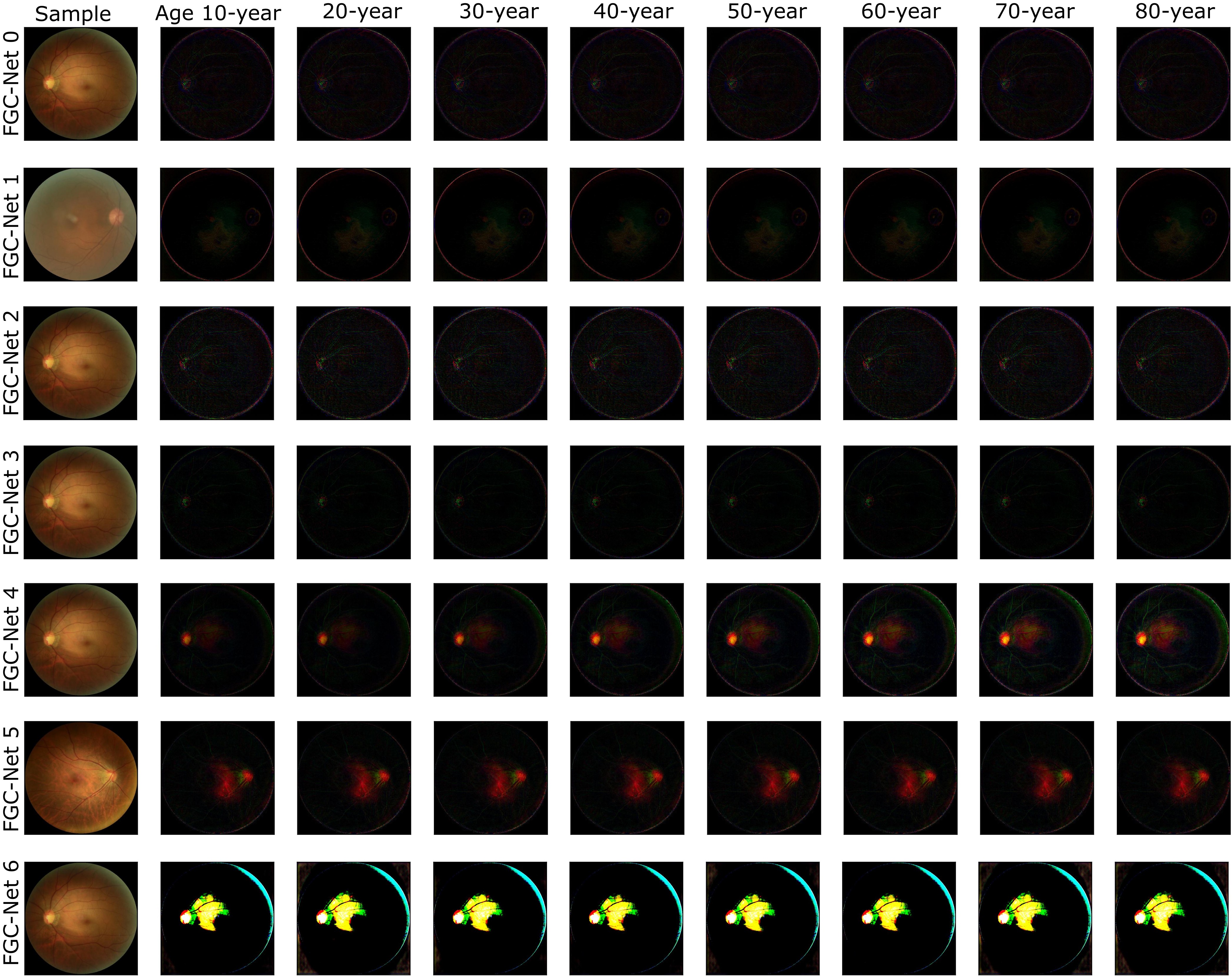}
	\caption{Displaying output results from distinct versions of FGC-Net. Each row corresponds to the FGC-Net version (from FGC-Net 0 to FGC-Net 6). There is total 9 columns, where first column displays randomly chosen sample. Second to ninth column shows the subtracted results between the input (sample) and corresponding output with the given condition. }
	\label{fig:figure4-fundus-images-generation}
\end{figure}

In this study, FAG-Net has been utilized to estimate biological traits from fundus images. After the successful estimation of age (accuracy 91.878\%) and gender (MAE 1.634), we proposed FGC-Net, a generative model conditioned by subjects' age. To extrapolate the effects of the fed condition, we proposed different versions of FGC-Net in order to verify the changes made on fundus images with age progression. After training FGC-Net (Figure~\ref{fig:figure3}) together with all versions, we randomly chose samples and fed them to the models to retrieve fundus images in different age stages (Figure~\ref{fig:figure4-fundus-images-generation}). There are total of seven versions of FGC-Net together with their output (Figure~\ref{fig:figure4-fundus-images-generation}). The random chosen sample is fed to each model together with different conditions (labels) as in the range of 10 to 80 years. The output images are subtracted from the original (fed fundus image) and the difference is displayed in Figure~\ref{fig:figure4-fundus-images-generation} (2nd column to the 9th).

Variations have been observed based on the subjective evaluation. From the visualized results, three key anatomical structures including optical disc, area near OD, and size (volume), were observed to be variant given different ages. The OD region, approximately a circular and bright yellow object, in all the generated fundus images by a variety of modalities found variant from early to late aging. The embedded age as condition mostly influences optic disc with age progression and which can be observed with naked eyes from 5th to 7th row (Figure~\ref{fig:figure4-fundus-images-generation}) for the corresponding model. Similarly, the nearby thick vessels and region to OD have also been observed variant with aging. Besides, the size of the fundus images has also been found variant with age progression. Such variations are apparent for FGC-Net-6 model (Figure~\ref{fig:figure4-fundus-images-generation}-last row). We employed attention mechanism in all the proposed models to highlight the regions of interest while embedding and estimating biological traits. The attention mechanism also highlights pixels in the input image based on their contributions to the final evaluation. Therefore, the affected regions in the generated images can be observed from the underlying modalities. The learning process from the embedded age as condition occurs at abstract level. In other sense, the learning becomes generalized by utilizing the fundus images from both healthy and unhealthy subjects and avoids. Thus, the study innovatively learns biological traits and their effects on fundus images using the cutting-edge technology of deep learning.  
The ability of neural networks to use greater abstractions and tighter integrations comes at the cost of lower interpretability. Saliency maps, also called heat maps or attention maps, are common model explanation tools used to visualize model thinking by indicating areas of local morphological changes within fundus photographs that carry more weight in modifying network predictions. 
Algorithms mainly used the features of the optic disc for gender prediction, which is consistent with the observations made by Poplin~\cite{poplin2018prediction}. Deep learning models that were trained using images from the UK Biobank and EyePACS data sets primarily highlighted the optic disc, retinal vessels, and macula when soft attention heat maps were applied, although there appeared to be a weak signal distributed throughout the retina~\cite{poplin2018prediction}.

\section{Conclusion and Future Directions} In this study, we investigate biological traits from fundus images from both healthy and unhealthy subjects. We also extrapolate the variational effects on fundus images with age progression. We proposed two types of DL models name FAG-Net and FGC-Net. FAG-Net estimates age and classifies subjects from fundus images utilizing the dense network architecture together with attention mechanism at distinct levels. The proposed models generalize the learning process in order to avoid the variation in anatomical structure in fundus images caused by the retinal disease. The study successfully carried out age prediction and gender classification with significant accuracy. Similarly, the attention mechanism highlighted regions of interest that are vulnerable to aging. Furthermore, the model shows similar salient regions in ungradable input images as in gradables (Fig. 2). This suggests that the model is sensitive to signals in poor quality images from subtle pixel-level luminance variations, which are likely indifferentiable to humans. This finding underscores the promising ability of deep neural networks to utilize salient features in medical imaging which may remain hidden to human experts. In the future study, more sophisticated deep learning models with attention mechanisms can be proposed for healthy and unhealthy subjects both in isolated and joint form.

%\bibliography{Fundus_images}
\bibliographystyle{IEEEtran}

\end{document}